# Transreality puzzle as new genres of entertainment technology.

**Ilya V Osipov**
Human Computer Interaction Laboratory, I2I STUDY, INC
San Francisco, CA
ilya@osipov.ru

## TAGs:

Puzzles; Mixed reality; Gamification; Tangible user interface; Interactive art and entertainment;   HCI

## Abstract

The author considers a class of  mechatronic puzzles falling in the "mixed-reality" category, present examples of such devices, and propose a way to categorize them. Close relationships of such devices with the Tangible User Interface are described. The device designed by the author as an illustration of a mixed reality puzzle is presented.

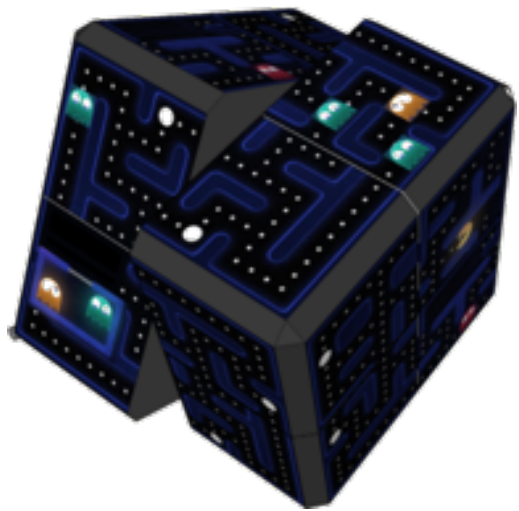

2017

# Transreality puzzle as new genres of entertainment technology.

## Introduction

Wide application of microelectronics and ubiquitous computerization leads to a situation, where computers are present almost everywhere. They are built into all things, and we are surrounded by them (pervasive computing) [ Weiser, 1991 ].The concept of the Internet of Things means that microcontrollers are built into all household items, from door locks to hair dryers, linking them in a common network. At the same time, a continuous decrease in the price of microelectronic appliances and children's interest in all things electronic give birth to brand-new toys and construction sets using built-in microprocessors. Due to the invasion of cell phones to overall life, gaming and computing are everywhere serving for entertainment, gamification (as a motivation tool driving this or that behavior model), education, and serious games.

Pervasive gaming as a variety of pervasive computing is the gaming process, which mixes the game universe with the physical world. The term itself is connected with ubiquitous games, augmented- and mixed-reality games, transreality and affective gaming, virtual reality games, smart toys, location-based or location-aware games, and crossmedia games [ Nieuwdorp., 2007] [Montola, Markus et al., 2009].

Mixed- or hybrid-reality gaming is the process, which takes place in actual reality and virtual reality simultaneously [Bonsignore et al., 2012]. According to Souza de Silva and Sutko, the formal characteristic of such games is "the absence of the primary game space; such games are played simultaneously in physical, digital, or represented spaces (such as the playing field)" [de Souza e Silva, et al 2009]. Allowing for the mixed-reality concept, which was proposed by Paul Milgram and is presented in Fig. 1 [ Milgram, et al, 1994], and the "virtuality continuum" (Fig. 2), the list of such games includes the popular game Pokemon GO, where players, while traveling in physical reality, affect what is happening in virtual universe, and what happens in virtuality motivates the players to perform physical moves in the real world. Other examples are The Killer, The Beast, Shelby Logan's Run, BotFighters, Mystery on Fifth Avenue, Momentum, Pac-Manhattan, Epidemic Menace, Insectopia, Vem Gråter, REXplorer, Uncle Roy All Around You and Amazing Race, as well as the transreality puzzles considered herein.

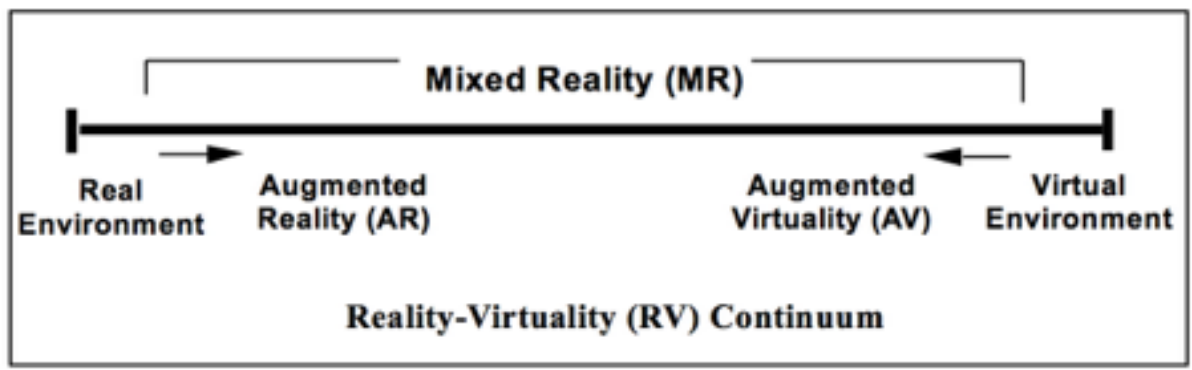

**Fig. 1. " Simplified representation of a RV Continuum." From Poul Milgram, [ Milgram, et al, 1994]**

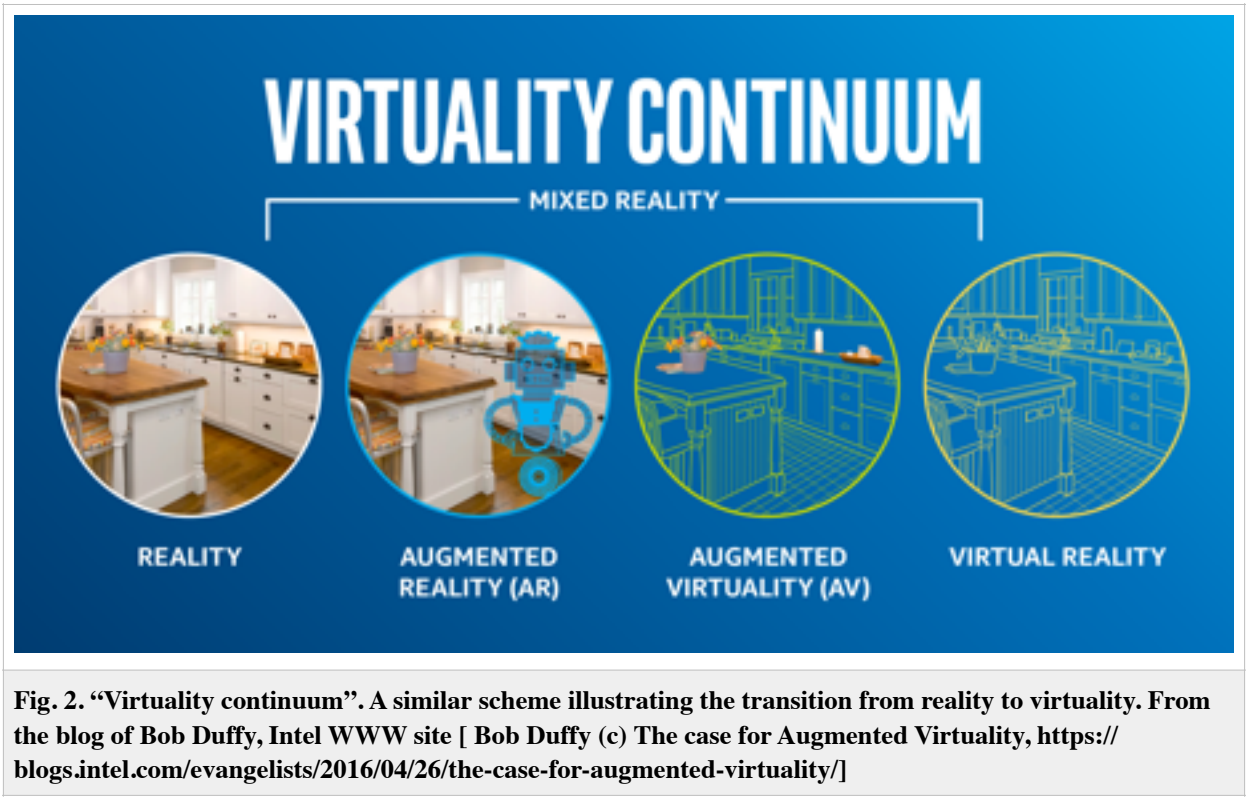

**Fig. 2. "Virtuality continuum". A similar scheme illustrating the transition from reality to virtuality. From the blog of Bob Duffy, Intel WWW site [ Bob Duffy (c) The case for Augmented Virtuality, https://blogs.intel.com/evangelists/2016/04/26/the-case-for-augmented-virtuality/]**

Augmented- or mixed-reality games, which are available now, are usually based on two techniques. The main technique is that the game processes the video camera signal and superimposes additional elements on the image of the real environment. (For example, the 2004 cell phone game called Mosquitos displayed the image taken by the phone camera on the screen of the phone, and overlaid images of giant mosquitoes on it, which the player had to shoot at using using the superimposed crosshairs). The second technique uses geolocation to combine virtual objects and geography of the real world. By the end of 2016, multiplayer Pokémon Go had spread widely and become extremely popular. It uses the both techniques, namely, it superimposes virtual objects on real camera images and like events and objects to the real world map using geolocation.

However, a whole class of games or gaming devices stay frequently beyond consideration. It comprises devices or sets of devices, which interact with the user physically, via positioning, slanting, or turning of their elements, thus affecting what is happening in virtual space, and this influence is often direct, since virtual objects are correlated with physical ones. The author proposes that such devices should be called "transreality puzzles". By their nature, they belong to the Mixed Reality realm ( Fig. 1), and, according to the scale proposed by Paul Milgram, are located in the Augmented Reality zone (AR). In what follows, the author reviews such devices

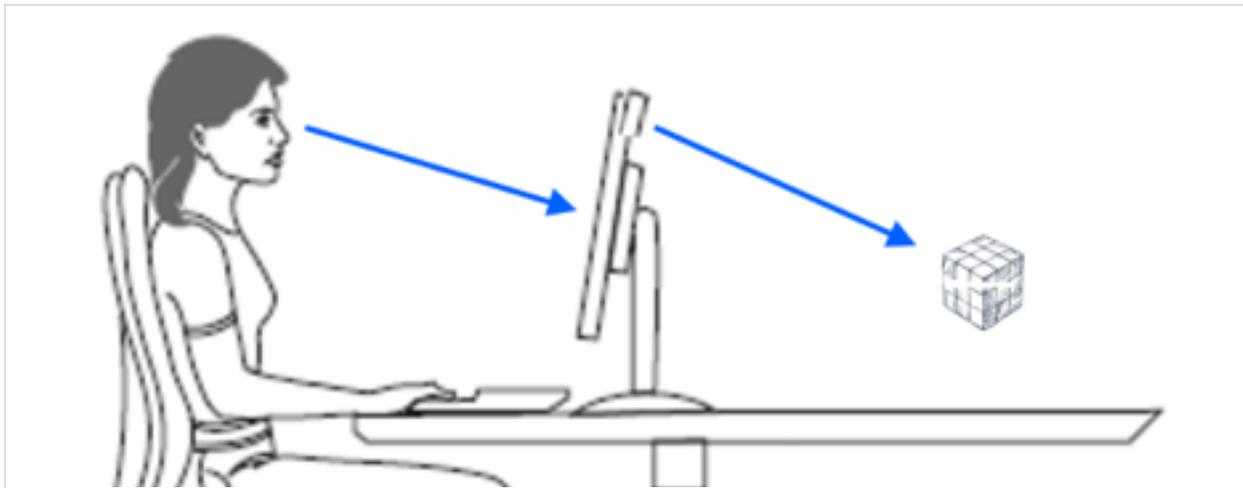

**Fig. 3. AR Puzzle. Scheme of working with a hypothetic puzzle in augmented reality. The user observes a physically existing puzzle (a Rubik's cube with additional icons on its sides) on a computer monitor.**

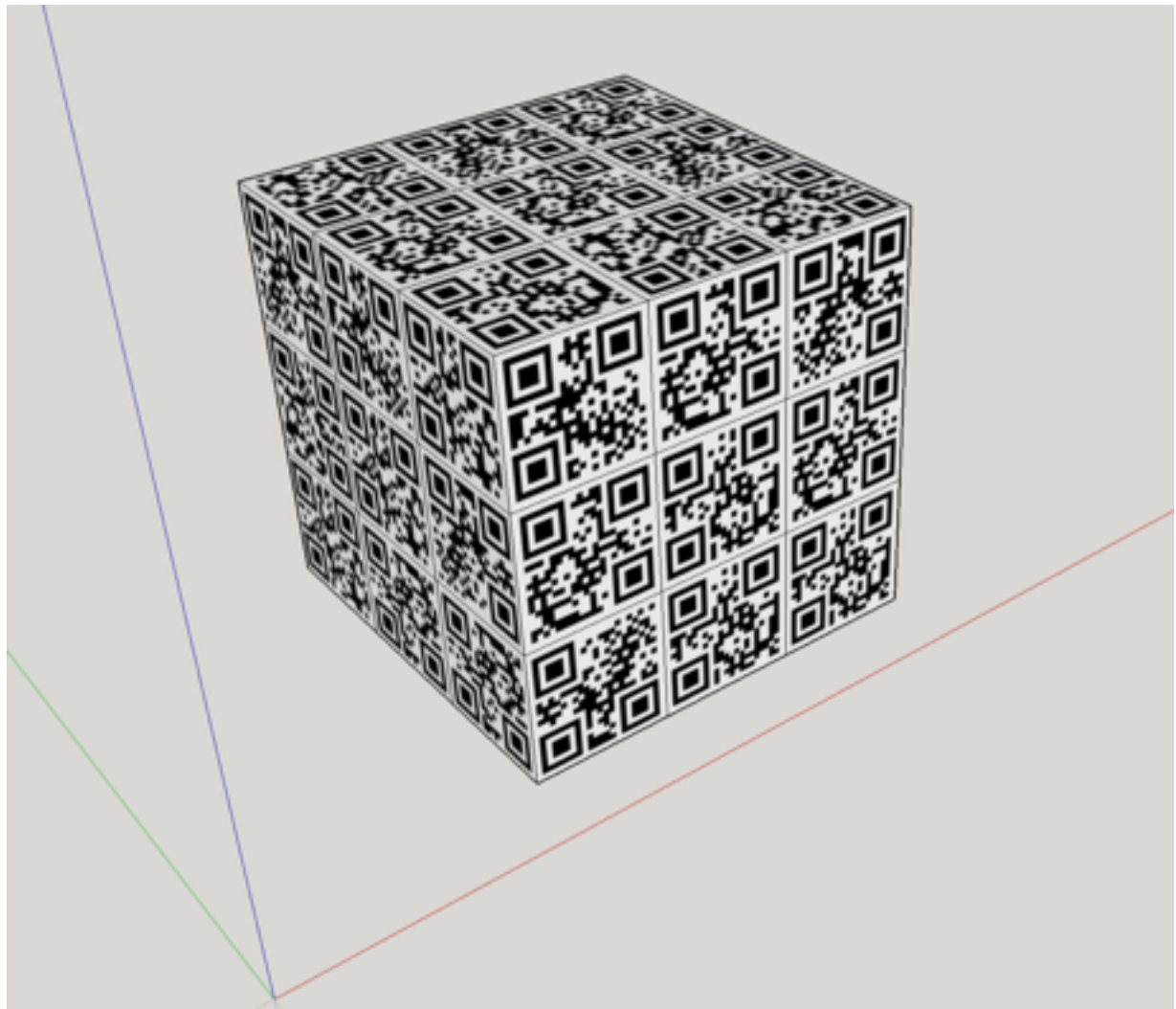

**Рис 3.1. QR-Code on each side of the AR puzzle.**

and projects.

J. van Kokswijk considers the "interreality system" as a system of virtual reality combined with real details, where the devices and physics that comprise it exist in the both realities under consideration [Kokswijk, et al, 2003]

The authors of [Gintautas, Hübler, 2007] describe an interreality pendulum including a real physical pendulum connected with its virtual counterpart. This system has two stable states: "double reality", where the motions of the pendulums are uncorrelated, and "mixed reality", where the pendulums are phase-correlated.

The author believes that interaction with a computerized puzzle or a construction set falls under the category of "transreality gaming" or a similar "mixed-reality games" category, where the gameplay take place simultaneously in the virtual (computer) medium and in the real world. The both aspects of the gameplay are interrelated and affect the both sides. The term "transreality gaming" was introduced by Craig A. Lindley [Lindley,2004] [Bonsignore, et al., 2012].

Figure 3 shows a hypothetic puzzle from the Augmented Reality category. A physical puzzle (e.g., a Rubik's cube) can bear marks (e.g., QR codes on the cube sides in Fig. 3.1), which are perceived by the computer and replaced by animated images and/or game characters.

*(A similar project called inSide was developed by the Tangible Media Group (MIT) for observation of the internal structure of three-dimensional objects. The AR effect is achieved by using projectors* [ Tang, Sekikawa, et al., 2014].)

However, an equivalent result can be achieved by installing displays directly on the sides of the puzzle. They will show virtual objects, layers, or characters. Such an object equipped with displays will be easier to hold by hand and observe, as compared with looking at it image on a display or by means of special glassed (see Figs. 18 and 19). Such a device, though it is equivalent to the above-mentioned system, can hardly be classified as an augmented-reality object, since it contradicts the accepted notion of augmented reality, which supposes that the user observes objects on a display or using a special tool: "Augmented Reality [is a] form of virtual reality where the participant's head-mounted display is transparent, allowing a clear view of the real world" [Milgram, et al., 1995]

From the taxonomy viewpoint, allowing for the fact that such devices do not fulfill the AR requirements completely, while still satisfying the conditions of a wider notion of Mixed Reality and being specific gaming gadgets of the Transreality Gaming category [Lindley, 2005] [Montola, 2005], they can be categorized as Transreality puzzles.

At the same time, such puzzles have all characteristics of TUI (Tangible User Interface) and can be considered as "Grasp Interaction" objects.

## Puzzles. Connection between game mechanics and interfaces

Combining puzzles and video games with the principles underlying the development of tangible interfaces reveals a wide range of new possibilities greeting the users of such devices, in the context of both the gaming potential per se, and useful entertainment. They are fit to be used by children, adults, and seniors, and develop speculative powers and three-dimensional skills.

The author regards puzzles as a kind of tactile mechanical representation of mathematical problems. As a pastime or a game, puzzle solving dates back to most ancient times. The game is a global phenomenon existing in all historical and modern cultures. Kant called gaming "purposiveness without a purpose (Zweckmäßigkeit ohne Zweck) [Kant, 1787, p. 136]. Original logical problems are found on the walls of Egyptian pyramids, in Ancient Greek inscriptions, and other historical records. In mediaeval history, the end of the IX century can be regarded as the golden age of puzzles. Puzzles became most common at the end of the XIX century and the beginning of the XX century. Owing to Sam Loyd in the USA and Henry Dudeney in Great Britain, puzzles invaded many periodicals and became popular among large sections of the public. For a long time, Loyd was thought to be the author of the 15-puzzle, which was a worldwide craze at the end of the XIX century (actually, it was invented by Noyes Palmer Chapman, a postmaster in Canastota, New York). The next milestone in the puzzle evolution was the invention of the famous Cube by the Hungarian engineer Ernő Rubik in 1974. Rubik's Cube was not only a toy, but also an object of studies of mathematics and engineers. Since that moment, "speedcubing" contests have been held regularly in the entire world.

**The most popular puzzles are:**

The **15-puzzle** (Game of Fifteen, Mystic Square) is a popular puzzle "invented" in 1878 by N. P. Chapman. It is a set of identical numbered square tiles in a square box, with one tile missing.

The **tangram** (Chinese: 七巧板; literally: "seven boards of skill") is a dissection puzzle consisting of seven flat shapes, which are put together to in a certain way to form a more complicated shape: a human being, an animal, a household item, a letter, a digit, etc. The shape to be produced is usually specified as a silhouette or an outline.

The **Tower of Hanoi** is one of the most popular puzzle of the XIX century. It consists of three rods and eight disks of different sizes which can slide onto any rod. The puzzle starts with the disks in a neat stack in ascending order of size on one rod, the smallest at the top, thus making a conical shape.

**Rubik's Cube** (originally called the **magic cube**, bűvös kocka in Hungarian) is a mechanical puzzle invented in 1974 and patented in 1975 by Ernő Rubik, a Hungarian sculptor and professor of architecture.

Puzzles can be regarded as the result of gamification of many problems, including those in the domains of education, development of spatial intelligence, and serious games. Ernő Rubik worked at the Department of Interior Design at the Academy of Applied Arts and Crafts in Budapest and built his cube as an illustration of the mathematical theory of groups.

Gamification is application of gaming practices in a non-gaming context. It is applied usually to motivate or demotivate a user to pursue this or that behavior model. This notion is also closely related to the concept of a "serious game", i.e. a game designed for some other primary purpose, rather than pure entertainment. The word "serious" is used generally to mark games used in such fields as education, defense, scientific studies, healthcare, emergency control, municipal planning, technology, and politics [Djaouti et al., 2011].

The connection between the tangible user interface (TUI) and success of gamification is dramatic. An appropriately chosen tangible interface can transform a serious game or gamification mechanics entirely. It can raise them to the heights inaccessible by using conventional methods, motivate the user to pass hours after hours of his or her time gladly, repeating the gaming experience, while reaching the specified educations purposes and ensuring the commercial success of creators of such devices, as well as their competitive advantages and widespread application of the solutions developed. Nearly all modern systems operated in mixed or hybrid reality employ TUI elements to a varying degree.

## Tangible User Interface (TUI)

For almost forty years, the keyboard and the mouse have been the main means of the human-computer interaction. They are supplemented by the graphical user interface called the WIMP model (windows, icons, mouse, pointer) [Bainbridge, 2004][Jacob, 2007]. Almost all TUI projects contain some of the innovations listed below: touchscreen, accelerometer, computer mouse, computer vision, or various sensors[Browne, Anand, 2012]. As they are adopted, simplified, and introduced into general practice, such interface devices and methods become mundane and start to be used in conventional graphical user interfaces.

The evolution of user interfaces started with the command-line interface invented in the 1960-70s. The first versions of the graphical interface were developed at the Xerox Palo Alto Research Center in 1970. Now, GUIs are used everywhere on personal computers and mobile devices. User interfaces are developed further as a series of interconnected designs and concepts called conventionally tangible user interface, kinetic user interface, natural user interface, touch user interface, organic user interface, natural language user interface, kinetic user interface, gesture recognition[Zappi, P., Milosevic, B., Farella, E., & Benini, L. (2009). Hidden Markov model based gesture recognition on low-cost, low-power tangible user interfaces. Entertainment Computing, 1(2), 75-84.], and even brain-computer interface[Liarokapis, et al., 2014].

The tangible user interface falls under the "tangible interaction" category. It is a variety of the user interface, in which a human being interacts with electronic devices by using material objects and structures [Ullmer, Ishii, 2000] [Shaer, Hornecker, 2010] [Hornecker, Buur, 2006].

Hiroshi Ishii, the most famous researcher of this field, creator and coauthor of dozens of devices, a founding father of this research line, says that TUI brings computing back 'into the real world' [Ishii, Ullmer,1997]

Donald Norman, the usability guru, states that "natural correspondence, i.e., the use of natural analogies and cultural standards, leads to instantaneous understanding" [Norman, 2013].

The advantages of tangible interfaces were first recognized by manufacturers of video games, specifically, electronic arcade machines. Special TUI-style devices were used in such machines already in the 1980s. Toy guns, steering wheels, pedals, joysticks provide much better interaction with video games. Additionally, game manufacturers started to build sensors into their products. For example, Boktai for Game Boy Advance was equipped with an UV photo sensor, and real sunlight was required to charge the "weapon" in order to shoot vampires. Nintendo produced WarioWare: Twisted with a piezoelectric gyroscope. In addition to using buttons, players were supposed to shake, twist, and jiggle the console. Nonstandard controls imitating natural interaction with material subjects became a standard for Nintendo DS games: for example, in Zelda for DS, players should blow at the screen to propel the sails of windmills.

Gradually, TUI elements appear in unspecialized devices. Specifically, the iPhone interface is based partially on the principle of direct interaction with objects. "People do not understand that we invented a new class of interfaces," Steve Jobs said in 2007, and explained that "the matter is that the iPhone interface lacks verbs almost entirely." This peculiarity, which the head of Apple emphasized, is really very important. A user of the traditional graphical interface first chooses an object, and then chooses an action to perform from a menu (that very verb Jobs was talking about). A TUI user does not need a "verbal" menu: he or she picks an object and does what he or she needs to be done. This is how multitouch gestures understood by the Apple phone work, it also works with browsing and map viewing in iPhones, as well as with inertial scrolling [Markoff, 2007].

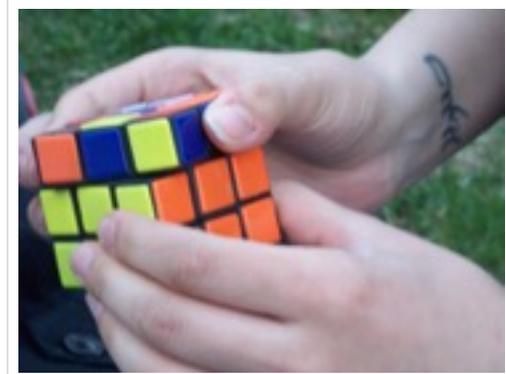

TUI techniques are also widely used in robotics, especially in those kinds of robots, which are designed to interact with humans [Guo, Sharlin, 2008]. Analysis of TUI interfaces, which can potentially be used for the purposes of gamification and serious gaming, is given in the review [Osipov, 2016].

**Fig. 4.** Rubik's Cube held by a user. Photo by Jessica Rossi https://www.flickr.com/photos/jesswebb/2586143138 licensed by creativecommons.org/licenses/by-sa/2.0/

## Grasp Interaction.

Puzzles and construction sets use the grasping effect brilliantly. This effect is per se a highly expressive input channel for interaction of a human being with material objects. It is always present, when we interact with small objects, and it is bidirectional, since the object provides a direct tactile feedback to a user's hands.

Grasping enhances motivation and ease of using multiple various artefacts. A good example to illustrate the grasping potential is the cell phone, which is so popular because one can grasp it, twist it, and hold it in hands. Recently, grasping has become the object of research performed by the human-computer interface community [Wimmer, 2015] [Johnson-Frey, 2003].

## Transreality puzzles, similar devices and construction sets

In this section, the author will review puzzles and construction sets falling under the category of augmented reality games and, at the same time, serving as the elements of TUI interfaces.

The review does not include games based on jigsaw puzzles (dissection puzzles, where the player has to put together a set of fragments of variously shaped images), e.g., Ravensburger Augmented Reality Puzzle. Puzzle video games, e.g., Minesweeper, are omitted from consideration also, since the author regards them as falling under a different game category of a similar name.

Transreality puzzles are electronic-mechanical puzzles (usually three-dimensional ones), where the gameplay takes place both in the domain of a virtual game, and in that of a mechanical device.

1. **Frazer's Intelligent Beermats** (1980) and **Threedimensional Modelling System** (1980) were among the first cubes, where the physical shape was augmented by an electronic game (see Fig. 5). From the author's viewpoint, they approach the mixed reality category, but do not enter it, since the virtual space of the games is manifested weakly.

This is what Kelly Heaton, the Physical Pixels developer, tells about this project in her doctorate theses:

"John Frazer and his colleagues at the London Architectural Association have been innovating alternate representations of computation since the late 1960s. For thirty years, Frazer's research has been motivated by an "evolutionary architecture", or one in which the computer contributes generative rules, responsive behavior and other life-like qualities to building structures. Not surprisingly, he finds the two-dimensionality of monitors and plotters restrictive for his exploration of a three-dimensional material space, and his background in architecture naturally leads him to physical modeling systems" [Heaton, 2000]. (Fig. 5)

## 2. TangledCube

Within the framework of this project, designers from the Faculty of Environment and Information Studies, Keio University (Japan) created a Rubik's Cube, which serves as a computer interface and displays positions of its sides on a computer monitor in real time. The virtual space of the game is represented only on the display, which turns TangledCube into the interface of an entirely virtual (VR) game, rather than a self-sufficient instrument [Kamada, Kakehi (2010)]. (Fig. 7)

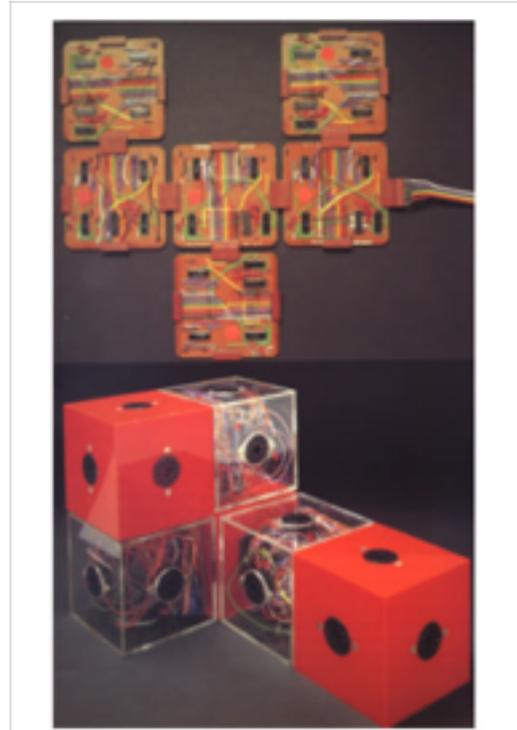

**Fig, 5.** Frazer's Intelligent Beermats (1980) and Threedimensional Modelling System (1980). (From Frazer, J. An Evolutionary Architecture. Architectural Association: London. [Frazer,1995] )

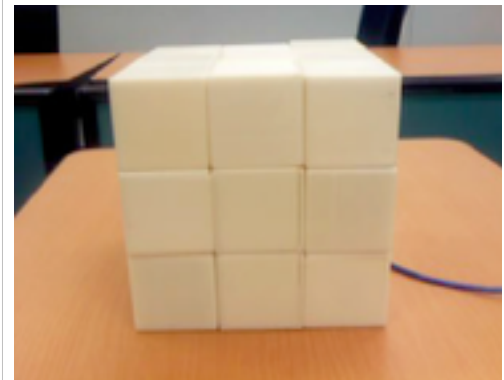

**Fig 7. TangledCube** (Photo from TangledCube: A Proposal on an Electronic Puzzle Using a Rubik's Cube-typed Tangible Interface Yohei Kamada, Yasuaki Kakehi)

### 3. Triangles

This construction set was designed at the MIT Media Laboratory. It allows one to use triangles to build sophisticated shapes and review the result displayed on a computer monitor. The physical computer interface is a construction set consisting of identical plane plastic triangles, which are joined both physically and digitally, by means of magnetic conductive junctions. As in the previous case, the virtual space of the game is displayed on a computer monitor, which also makes Triangles an interface to an entirely virtual (VR) game, rather than a self-sufficient instrument [Gorbet, Orth, Ishii, 1998]. (Fig. 8)

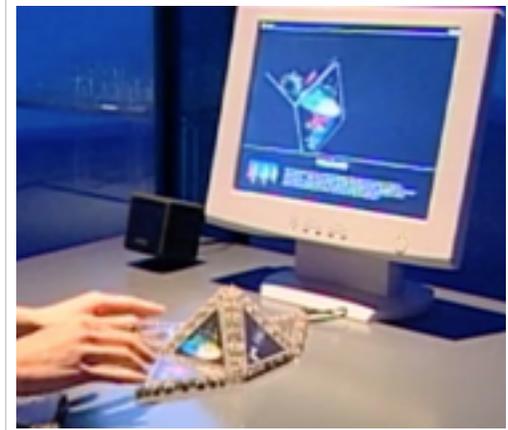

**Fig 8. Triangles** ( Photo from web-site MIT Media Lab http://tangible.media.mit.edu/project/triangles/)

### 4. ActiveCube

This project was developed at Osaka University. Cubes can assembled to form any shape, similar to LEGO. However, since each cube is equipped with a microprocessor and a parallel data transmission interface, the block of cubes "understands" the current configuration in real time. A computer receives the information about the collective set of blocks via the main block connector and shows the entire group on its display. If the operator moves blocks, the display shows the final configuration in real time. This project is similar to Triangles. The two projects, however, were designed by two design groups independently [Kitamura, et al., 2000]. (Fig. 9)

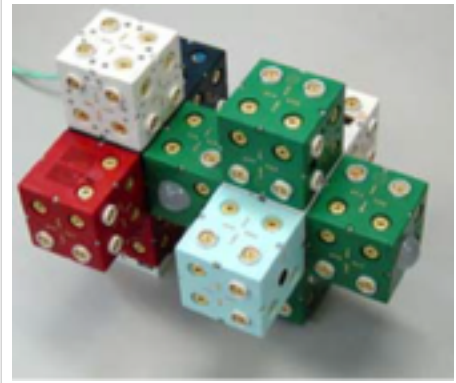

**Fig 9. ActiveCube** ( Photo from Watanabe [Watanabe, et al 2004])

### 5. Peano

Peano is a set, which consists of fifty cubes, which, when joined, form a modular full-color 3D display.

Each cube made of colorless diffuse plastic has an LED installed at its center. The LED illuminated the entire cube volume, and its color may vary. The cubes are combined in a modular network, which supports direct and software control of the animation of the structure color. Although the network topology is linear, its geometry is three-dimensional, which is possible due to the use of the Peano curve. This curve (as well as its variant, the Hilbert curve) is a linear structure turning at an angle of 90 degrees to determine three-dimensional space with Cartesian coordinates. Magnetic mechanical connectors are specially designed to maintain connection in four directions and form a linear peer-to-peer network. Peano cubes can be animated by using a digital palette. Since the cubes are sensory and "are aware" of the network topology, they can be controlled by touching them [Heaton, 2000]. (Fig. 10)

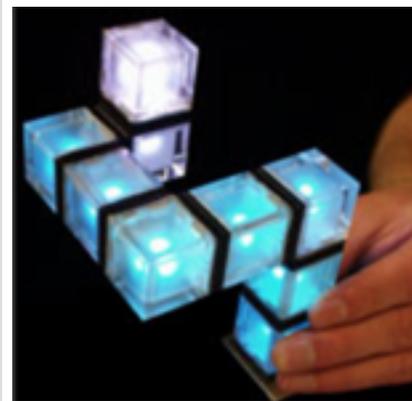

**Fig 10. Peano** ( Photo from Heaton, K. [ Heaton, 2000] );

### 6. Constructed Narratives

This construction set consists of 3D Pentomino tiles, which interact with each other and a PC via wireless channels and allow several people to play various games simultaneously. The PC displays shows their virtual representation and additional information, e.g., colors and owners of the tiles, or letters for playing.

According to the P. L. Jennings, the author of the project, Constructed Narratives is the first experiment in the field of studies of social tangible interfaces [Jennings, 2009] [Jennings,2005a] [Jennings,2005b]. (Fig. 11)

### 7. Siftables / Sifteo

In the author's opinion, this is the first full-scale device falling under the **transreality puzzles** category. The elements of Siftables (Sifteo is the commercial name of this project) are both physical objects and virtual game elements at the same time. No external computers or displays are required. Thus, this is the first construction set or puzzle, which is a mixed reality object.

Siftables elements look as toy bricks. The top of an element is a small-size LED display. Inside each tile, an accelerometer, four IR sensors (one per each side), and a microprocessor are installed. The tiles interact with each other and a PC. By means of the sensors, they "perceive" contacts with other tiles and can "feel", when they are lifted, shaken, or tilted. Test applications developed by the Siftables inventors at the MIT Media Laboratory demonstrate the familiar set of gestures used in tangible interfaces. A set of such tiles can be used as a platform for child's games [Merrill, et al. 2011] [Hunter, 2010]. (Fig. 12)

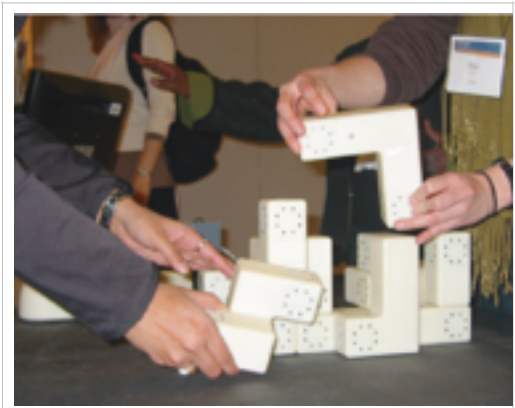

**Fig. 11. Constructed Narratives** (From the Web site of the project author , P. L. Jennings)

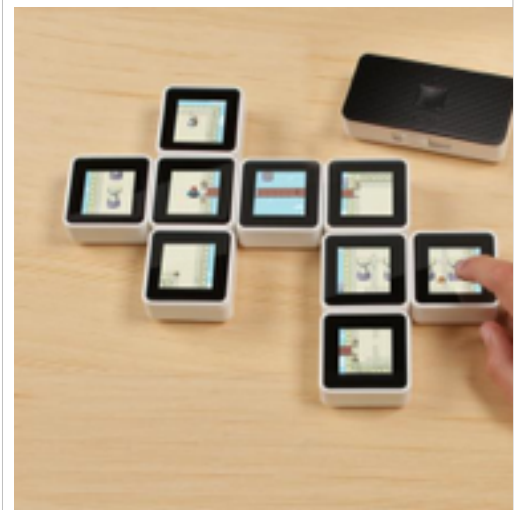

**Fig 12. Sifteo - photo from the Web site of the Sifteo project** (both the project and the site were closed at the date when the paper was written)

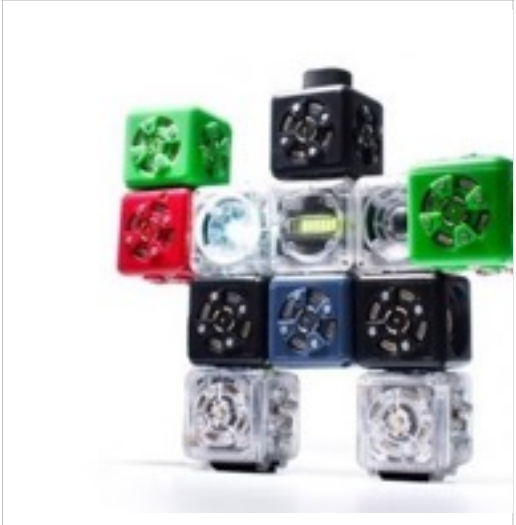

**Pic 13. Photo from the Web site of Modular Roboticss** ( http://www.modrobotics.com/cubelets/ )

**8. Cubelets**

The Cubelets game was developed by Modular Robotis. It is a toy set replaying the functionality of the ActiveCube and Triangles projects. However, its elements can be equipped with motorized units, as well as with sensors and other additional equipment. All this allows players to build toy robots, cars, and other items making the game more entertaining.

From the author's viewpoint, this project has few features in common with transreality puzzles. Rather, it is a computerized and motorized construction set and presented here for the sake of illustration only [Schweikardt, 2011]. (Fig. 13)

# TR-puzzle Cubios

The author has designed and made a prototype of an entirely "transreal" puzzle: a set called Cubios. The main purpose of this puzzle is to join physical representation and virtual gaming environment in a common indivisible gameplay scenario.

According to Paul Milgram's concept of mixed reality, which is illustrated in Fig. 1, the proposed puzzle will fit in the middle of the scale, between augmented reality (AR) and augmented virtuality( AV), since the gameplay will require elements of the physical and virtual world equally.

The gameplay lies at the boundary of the two media. Comparing it with the Siftables/Sifteo project, one can state that the latter is closer to augmented virtuality, since all Sifteo cubes can move around on a plane freely, and during the game, players focus mainly on what is happening in virtuality (see Fig. 14).

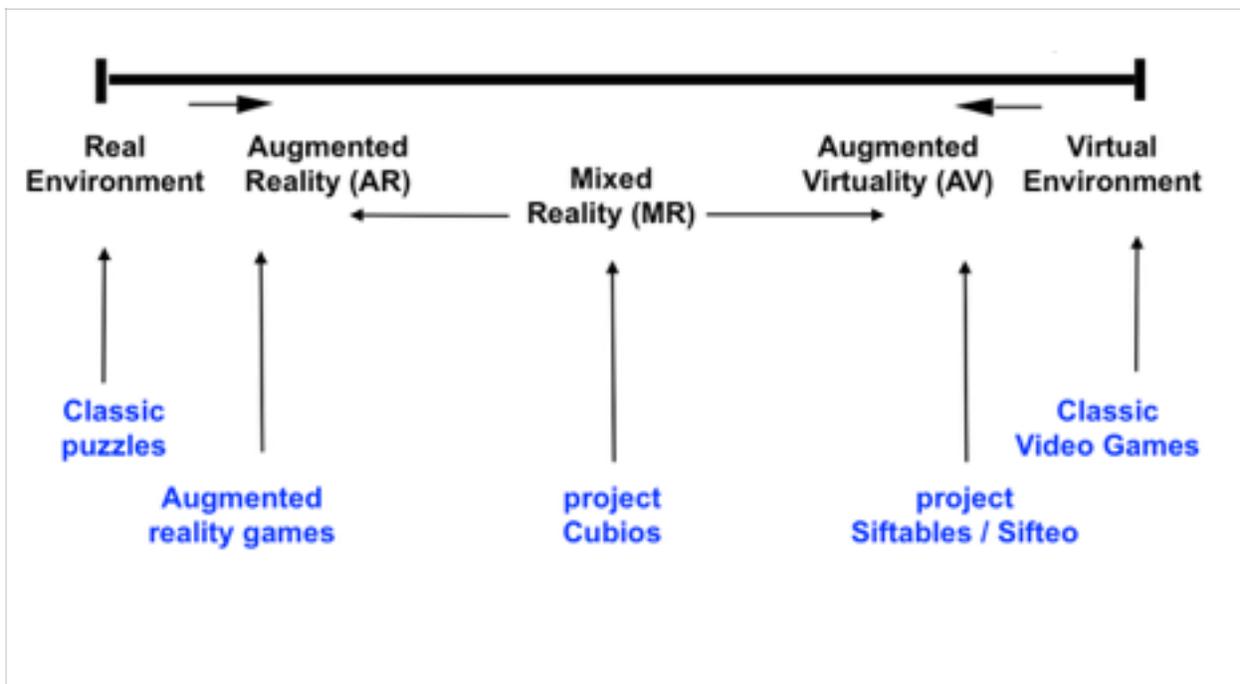

**Fig. 14. Positions of various games in the virtuality continuum.** Based on the simplified representation of the RV continuum by Paul Milgram and supplemented with positions of classical games, puzzles, augmented reality games, as well as, for the sake of illustration, those of the Siftables/Sifteo project and the Cubios puzzle.

Thus, Sifteo is closer to classical computer games, where 99% (nominally) of the gameplay take place in the virtual environment, and only 1% is left for the controls, such as a joystick, a mouse, a Kinect, etc. An assembled Cubios is supposed to be played as a twisty puzzle, where sides are to be moved according to specific rules, in order to arrive at the result required (e.g., gather all the tiles of the same color on one side in Rubik's Cube). In this case, the logic of the virtual game requires that parts of the puzzle should be moved and connected in a certain way, in order to achieve the result or advance in the gameplay. Thus, a significant part of the gameplay takes place in the physical world, and the other, not less significant, in virtuality.

The author proposes a special magnet-operated construction set designed for a gameplay in physical reality with an associated game taking place on built-in displays or LED matrices, in virtuality. Cubios, being equipped with microprocessors and displays, can be regarded as a variety of the Tangible User Interface, a transreality puzzle, and a gaming platform (i.e., a platform-specific combination of electronic components or computer hardware which, in conjunction with software, allows a video game to operate).

The idea is based on the 2x2x2 PocketCube puzzle invented by Larry D. Nichols in 1970 (US Patent 3,655,201, "puzzle and method with pieces rotatable in groups") [Nichols, 1972]. A special shape and design of each element (a "cubio") allows one to gather structures similar to Larry D. Nichols's puzzle and more sophisticated combination (see Fig. 15). The main difference between this design and the existing magnetic construction sets is the use of magnetic connectors based on magnetic balls (or other solids of revolution) and connectors on cubio sides for joining them with other cubios. Moreover, Cubios elements can be set on additional ball-shaped elements allowing one to specify pivot centers, or the mechanical puzzle Pocket Cube. As a result, the entire structure turns into a twisty puzzle [Sun, Zheng, 2015].

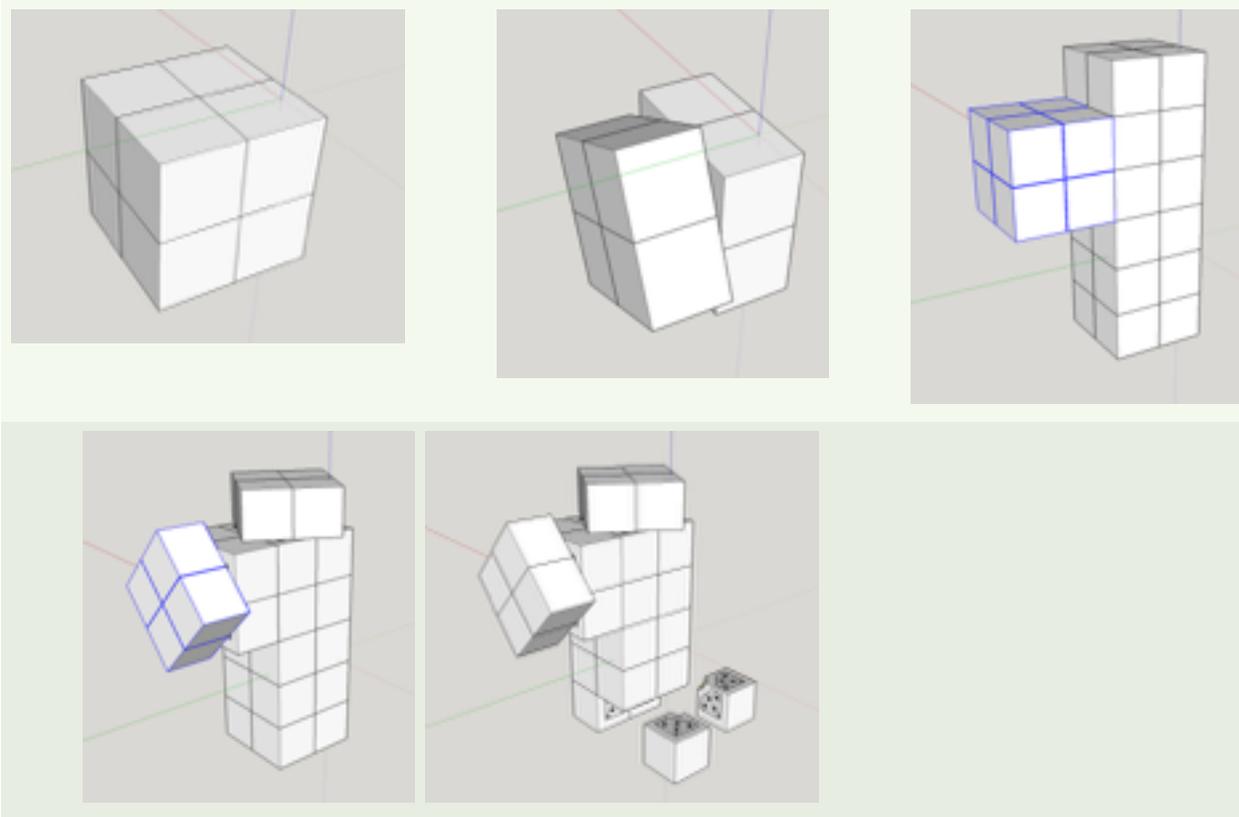

**Fig. 15. Variants of assembling Cubios**

To make a cubio satisfy the requirements of a transreality puzzle, it should be equipped with a microprocessor, and each of its external side should be a display (or LED matrix). All the microprocessors working together ensure cooperation of all cubios as an integral Cubios device.

Along with the standard configuration using eight cubios (2x2x2, as in Larry D. Nichols's patent), cubio tiles can be assembled to obtain other shaped shown in Fig. 15

 (to realize all the shapes shown in Fig. 15, one will have to use various cubios with different number of sides/displays and connectors).

**Cubios advantages:**

1. This is a new gaming environment, which can attract players due to its novelty and coordinate difference from arcade machines, computers, tables, cell phones, and TV consoles.

2. It develops spatial intelligence and, probably, cognitive skills of the player [Green, 2015] [Gray, 2015].

In its essence, Cubios is a superset in relation to Nichols's magnetic 2x2x2 Magic Puzzle [Nichols, 1972], which can be assembled of eight cubios.

Technically, many puzzles can be turned in transreality games. For example, if each tile in the well-known 15-Puzzle (Fig. 15) is a display or a screen, to which images are projected or superimposed for displaying by means of computer processing, then one obtains a similar device and gaming platform, where part of the gameplay resides in the real world, and the other part, in virutality. It can also be assembled on the basis of Cubios elements, by using 15 cubios (with displays on one side and connectors on four sides) and a box to house all these elements. Other shapes are also possible, e.g., a pyramid (requiring elements with triangular displays), a cube (3x3x3, 4x4x4, etc.), an octahedron, a hexagonal prism, a cylinder, a sphere, and so on.

Puzzle fans single out a special puzzle class, specifically, "twisty puzzles" (http://www.twistypuzzles.com/)( https://sites.google.com/site/geduldspiele/Gallery3DTwistyPuzzles) [Sun, Zheng, 2015]. They are puzzles with one or several axis of rotation, around which the main action takes place. Such puzzles are most popular (see Amazon: Best Sellers in 3-D Puzzles, https://www.amazon.com/Best-Sellers-Toys-Games-Puzzles/zgbs/toys-and-games/2491829011). Electronic puzzles, however, are rarely included in this category, since electronic appliances do not act as rightful drivers of motions along the axes of rotation. Rather, they are merely present or imitate their participation. The author found more than 20 popular electronic puzzles, but none of them fulfill the requirements of twisty puzzles entirely (see the list of electronic puzzles at http://twistypuzzles.com/cgi-bin/pdb-search.cgi?act=sec&key=24).

One of non-electronic analogies is the IM Puzzle Ball. It reminds Cubios in that it also based on Rubik'c Cube, which is augmented with a puzzle, where a ball moves in a maze (a mechanical game instead of the electronic one) (see http://www.jaapsch.net/puzzles/

impuzzle.htm , http://www.twistypuzzles.com/cgi-bin/puzzle.cgi?pkey=4287 ). However, this not an electronic puzzle, it lacks any virtual component and, thus, cannot be regarded as a mixed reality device.

The sides/displays of Cubios as a whole form a common gaming space, where game characters or elements interact with each other. The gameplay is controlled mainly, but not exclusively, by turning sides, similarly to Rubik's Cube gameplay, and/or mounting or dismounting individual cubios (even from a different set) and moving them afterwards. Additionally, the player can use the accelerometer, or display sensors, or any other sensor to control the gameplay. To create a clear feedback and increase the gameplay, tactile feedback can be used [Mazzoni, Bryan-Kinns, 2016]. However, the author believes that the main gameplay control method should be turning the sides of the assembled Cubios around their axes. This is the way to achieve balance between augmented reality and augmented virtuality, which makes Cubios a device allowing one to play a true mixed reality game.

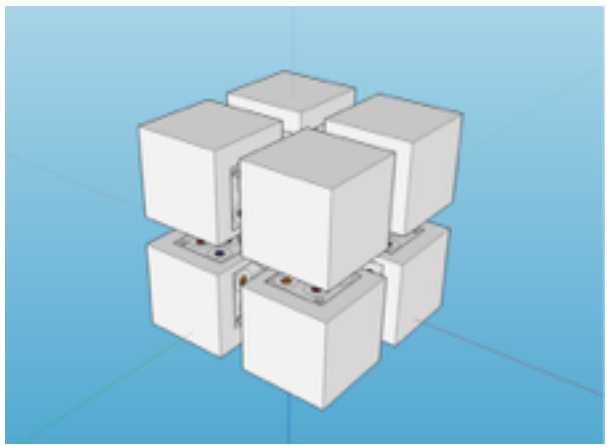

**Fig. 16a.** Standard configuration of Cubios using 8 cubio tiles. All tile connectors are directed inwards, all outer sides are displays (see Fig. 18). Such a structure can turn freely around 3 axes on the mechanical base in the center.

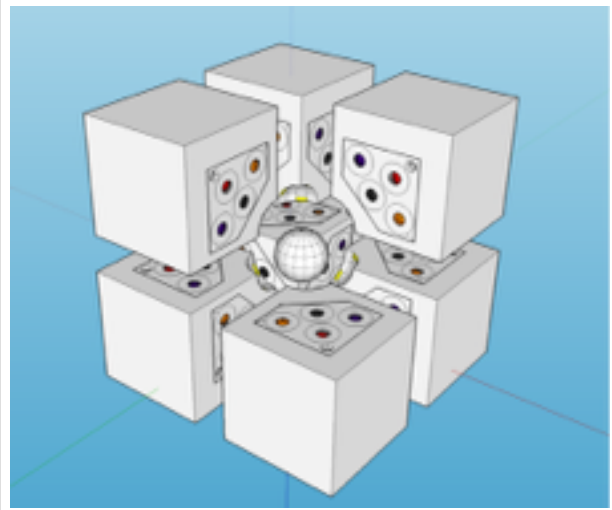

**Fig. 16b.** Same as in Fig. 16a with one cubio removed. The steel ball in the middle serves as the base, which allows the structure turn around 3 axes without bias. The ball can be replaced with a Pocket Cube mechanical puzzle.

The basis of a challenge to play the verge of moving to participate in an exciting game based on the characters, the puzzles, the selection of words, or just a "meditating" on a beautiful iridescent and interacting with each moving pictures on faces.

Ideas of some games, which can be developed using the Cubios set, follow:

1. **A variant of Scrabble.** Each of the 24 tiles is labeled with a letter (less frequent letters, e.g., q and z, are omitted). The game purpose is to turn Cubios sides and put letters together in such a way, as to make adjacent letters form a noun (or any word). The computer highlights the formed letter with a color, awards several points to the player, and suggests going on with the game. The player, who assembles more words than the other players out of random combinations during a specified period of time, wins.

2. **A variant of Pac-Man. The game runs on all six sides of the Cubios at the same time.** The player controls Pac-Man by tilting the device (using the accelerometer) and turning the sides of Cubios (see Figs. 18 and 19), trying to avoid ghosts and eat glowing dots, which make him a super-hero capable of eating the ghosts. If a ghost touches Pac-Man in the ordinary regime, the player loses. If Pac-Man eats all ghosts, the player wins and moves Pac-Man to the following level, where the game is played again at a higher speed and a higher degree of complexity**.**

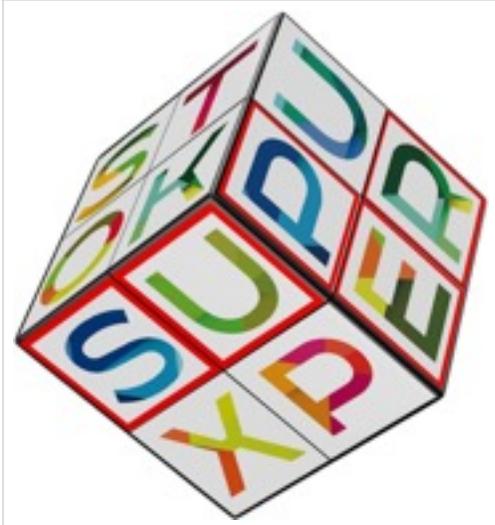

**Fig. 18. Cubios with the gameplay in process on all external displays forming common gaming space.**

3.**Maze.** A maze covers all sides of Cubios. The player should take the game character, who faces various dangers in the maze, from the entry point to the exit by turning the sides. This is an analogy of IM Puzzle Ball.

4.**Faces.** Fragments of six quarter-cut faces are shown on the Cubios sides in random combinations. The player should assemble the faces correctly. Move after move, thousands of possible combinations are formed (e.g., a portrait of a man with a mustache combined with that of a man wearing no mustache yields a temporary transitional face of a man with a pencil-thin mustache).

5. **Patterns/Fractals.** Each fragment is a different pattern, fractal, or, e.g., a floral ornament. When tiles are moved, ornaments "shoot out" to adjacent displays on the newly formed side, thus producing thousands of combinations. This game is purely aesthetic, there are no points, nor wins or losses.

6. **Bubbles**. Initially, each side shows bubbles or other geometric figures floating slowly in all directions. Turning a side "cuts" the bubbles which happen to be on two adjacent displays at the moment. Each part becomes a new bubble, smaller than the initial one. When touching each other, the bubbles merge, and their properties (color, texture, shape) are combined.

7. **Birds/Fish**. Birds fly or fish swim on each side bouncing off the ribs. The player can move characters by turning the Cubios sides. The purpose is to drive all birds or fish on one side.

8. **Pac-Man vs. Digger.** It is a mix-up with characters from two retro video games. Initially, they are put on different sides, but as the sides are turned, they pass from one side to another and start interacting.

9. **Cities and Landscapes.** At first, the puzzle sides show real photos of cities and/or landscapes, which are cut into fragments that can join seamlessly. When the sides turn, fragments of photos join in random, unanticipated combinations forming new amazing scenes.

10. **Mixing colors.** From the start, each Cubios side has its own color. By turning a side, we make colors mix gradually, rather than get tiles of several colors on one side, as on Rubik's Cube. This game can be useful for children as an illustration of color mixing. The same process can be played with various materials, for example, the "water" side, on mixing with the shiny "iron" side can make the latter "rust".

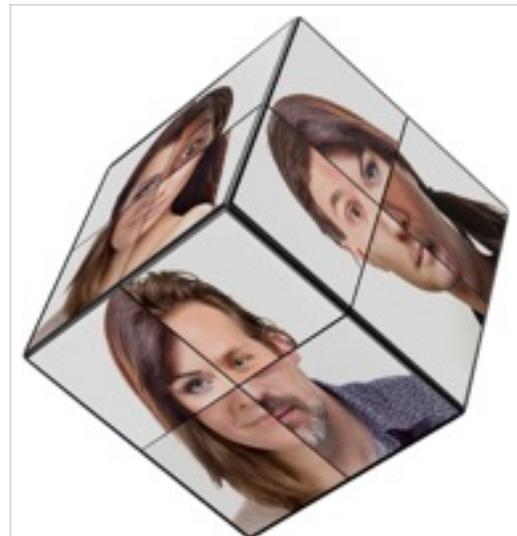

**Fig 17. Cubios with the gameplay in process on all external displays forming common gaming space.**

11. **Scissor-Paper-Rock**. The six sides show two "stones", two "papers", and two "scissors", one item per size. The images are centered. When a side is turned, the image changes. If half of the scissors happens to be on the same with half of the stone, the entire image turns to stone (stone wins). The game is over, when identical objects fill all the sides.

12. **12. Circles/Loopy Worms.** In the middle of each side, there is a circle with the center at the point where four cubios join. All six circles are of different sizes and colors. Turning of a side cuts circles into two halves, which turn slowly and appear on "foreign" sides. Turning a side again can cuts parts of the circles or arcs. The purpose of the game is to drive all six circles on one side.

13. **A version of 15-Puzzle,** only it is 23 now, rather than 15. One tile display is empty (all the other are labeled with numbers). When a side is turned, the tile adjacent to the empty one, replaces it. The purpose of the game is to arrange all the tiles in the numerical order (and, if they are colored, according to their colors as well). A similar game is virtual representation of the "Minus Cube" game, the three-dimensional version of the 15-Puzzle: seven cubes glued of two U-shaped halves, white and colored, can be moved in a transparent sealed box, which is twice as large as the cube. They should be arranged so that only one color could be seen on each side of the box [Zubryaev,1986].

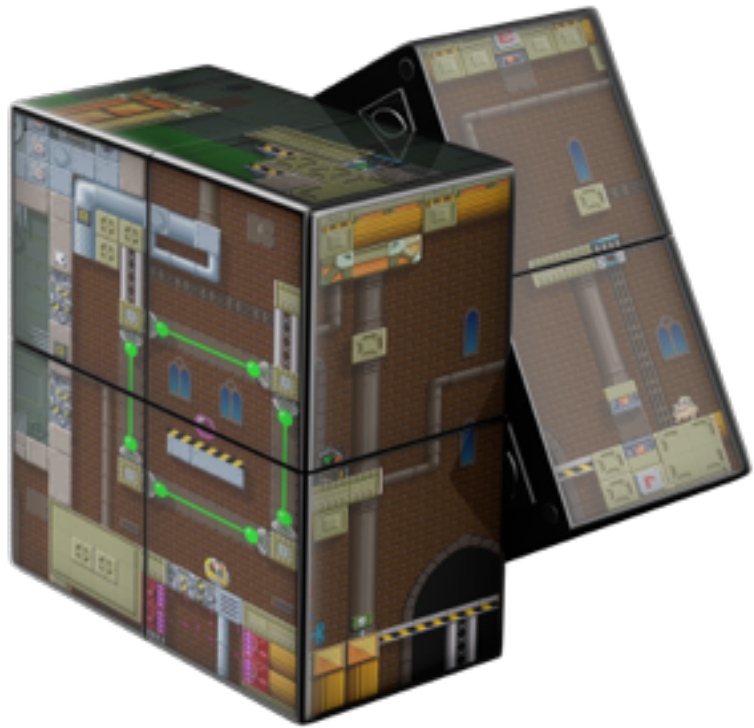

Fig 19. Cubios while turning around one of the axes.

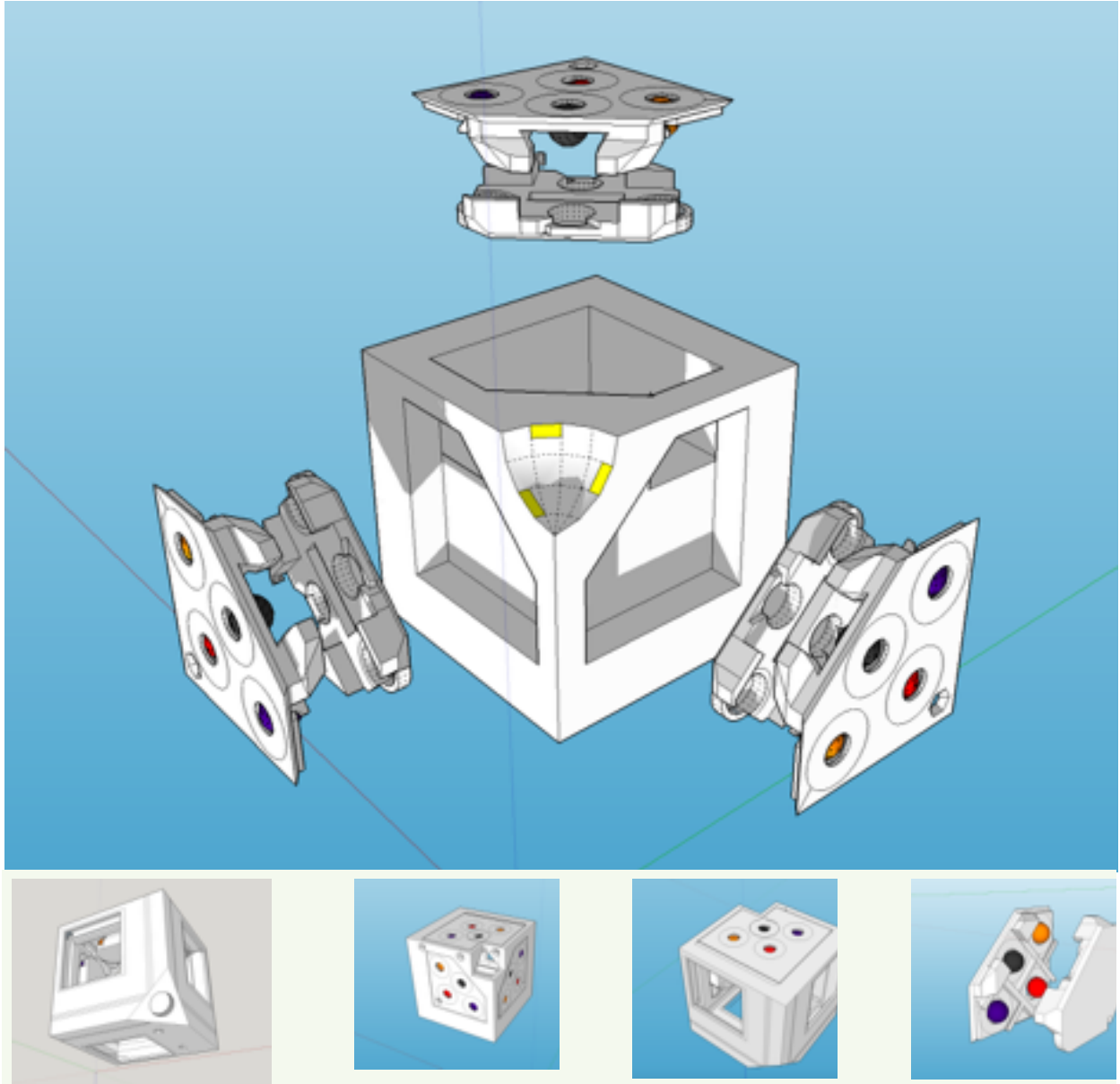

**Fig. 23. Fragments of the Cubios project.**

*Note: Since it is assumed that individual cubios are not connected rigidly, a player can rearrange them during gaming or add cubios from a different set. In this case, the gaming algorithm can treat this differently, specifically, accept this rearrangement as part of the game (if the scenario is not disrupted by this action), or reduce the score by awarding negative points or time penalties, regarding this trick as cheating, or even stop the gameplay whatsoever, taking this move as a defeat.*

The author has developed a prototype of the Cubios set. The software was written in the C Programming Language, electronics was assembled on the basis of an ATmega328 microprocessor, and cubio bodies and electromagnetic connectors were printed on a 3D printer (see Figs. 20 and 21).

An operable Cubios prototype is shown in Fig. 21. In future versions, displays will occupy all possible space on side surfaces, and the frames around them, as well as the gaps between them will be minimized.

**Important features of the design:**

1. Each cubio in the Cubios is an individual device equipped with all items necessary for independent operation, i.e., the battery, microprocessor and auxiliary electronics, and the charging connector.

2. Being held together magnetically, Cubios tiles start interacting in accordance with the gameplay scenario. Their microprocessors are synchronized, and the gaming process runs on several joint cubios as a common gaming field. The magnetic connections hold them together, but they are not rigid (by analogy with Larry D. Nichels's invention).

3. Permanent connection of the cubios is not required. They can be dismounted and mounted again during the gameplay..

4. The gameplay consists both in turning elements of the structure around several axes, and rearranging, mounting, or dismounting cubios (including additional ones from "foreign" sets).

5. Initially, all cubios forming the Cubios puzzle are equisignificant. They form a peer-to-peer network, and if a leading element is required by the gameplay scenario, it is chosen at random.

6. There are no mechanisms (other than magnetic contacts) in the puzzle. The game mechanics is determined only by the shape of the elements (cubios) and their magnetic properties. If rotation about some axes is required, a supporting ball or a Pocket Cube is placed in the center of the structure. It can be part of the structure (if the Pocket Cube mechanism is a structural element, the Cubios will be an undismountable unit).

7. No accelerometers or gyroscopes are required for individual cubios to perceive their positions in the structure. They "see" the current structure by queuing adjacent cubios and interacting with them.

**Technical description and patents**

In the process of working on the device 2 patents were submitted:

The first patent describes the operation of magnetic compounds whose main property is the neutrality of male / female connectors, as well as rotating self-orienting magnets, which allow to eliminate the problem of mutual repulsion of identical poles of N-N or S-S;

*EFS ID 27279988, US Application Number 62410786, Title of Invention: Magnetic electrical connector integrated in the flat surface of the product, with a spring effect and the possibility of rolling. 20OCT2016*

The second patent describes the design of the device as a whole:

*EFS ID 28446467, US Application Number 62462715, Title of Invention: Electronic Device With Display Surround Transformability. 23FEB2017*

The technical description of the assembled prototype is given in **Appendix 1.**

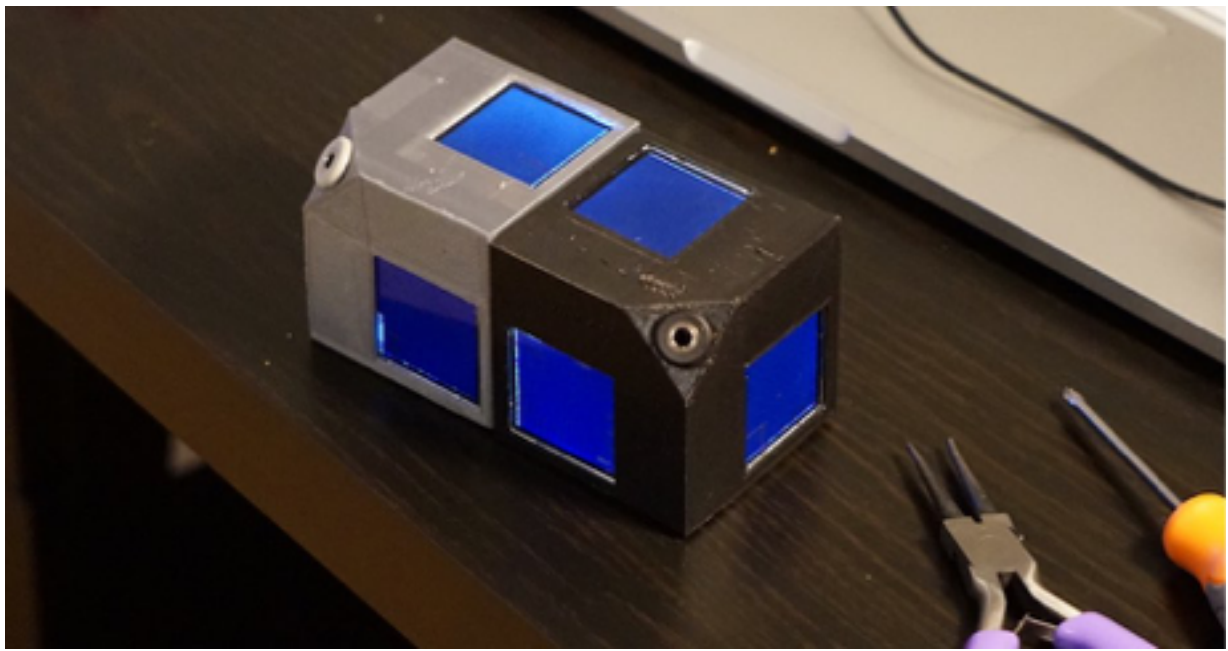

**Fig. 20. A pair of assembled cubios, joined together.**

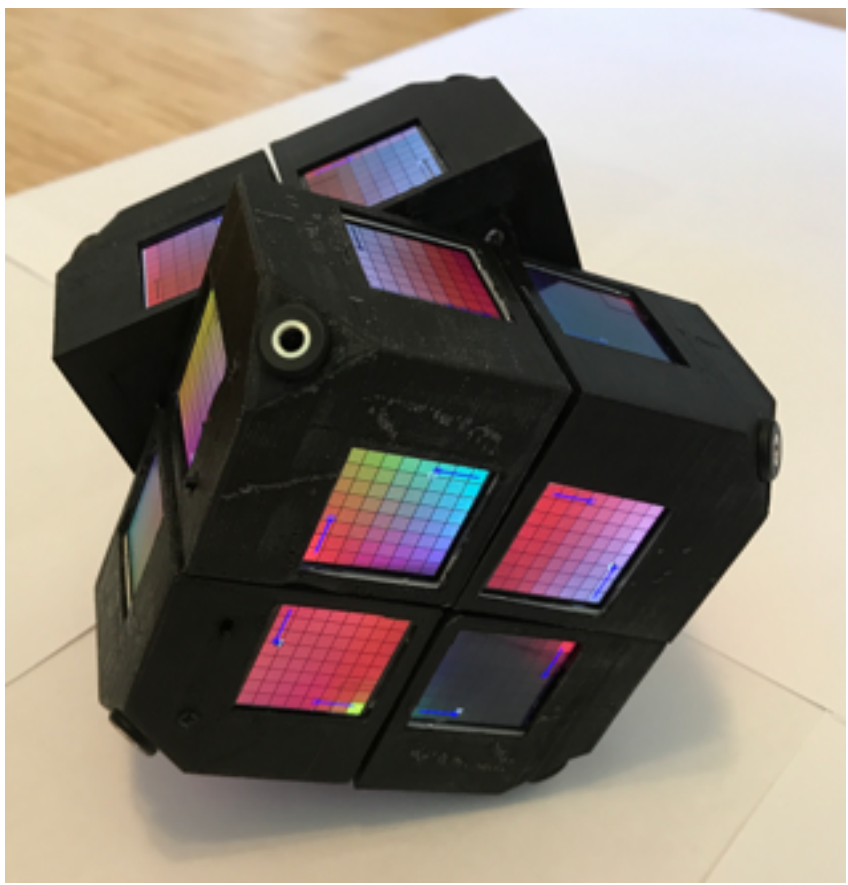

**Figure 22. The assembled and functional prototype Cubios of 8-elements**

## Discussion and plans:

The main question for such projects is how potential users will receive it. To find this out, it is planned to manufacture several full-scale prototypes, develop several full-scale games, and offer them to participants of test groups, who will play and study the device.

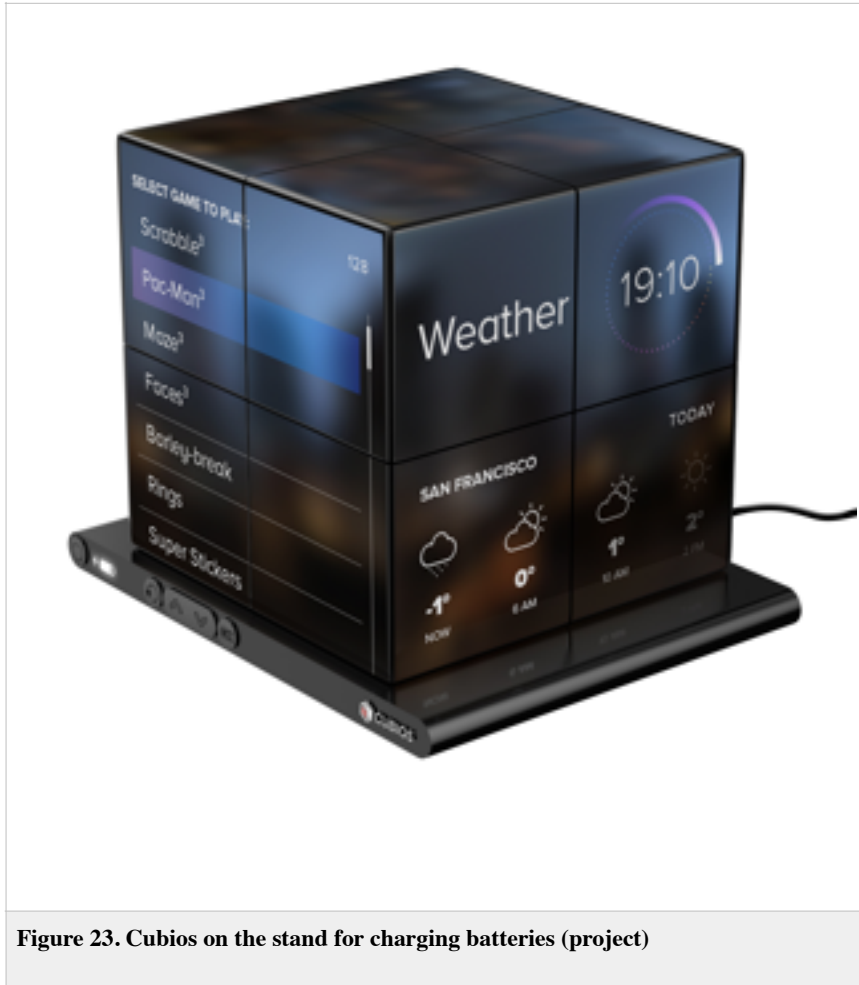

**Figure 23. Cubios on the stand for charging batteries (project)**

The test methods will be questionnaire surveys, test games, and contests (e.g., who will be the first to lead Pac-Man from one side to another without running into an obstacle, or who will be the first to make up ten words playing in the Scrabble regime). Test groups will include pre-school kids, schoolchildren, teenagers, as well as pensioners. Testing by representatives of puzzle fan communities would be most welcome, as well as running tests on random adults in the places, where people have to wait for a long times and would not object to being offered an interesting pastime, e.g., in airports.

In addition, using the focus group, it will be possible to determine, which type of the design (dismountabe or one-piece) is preferred by users.

The tests will be finalized by questionnaire surveys, both direct and comparative ones (i.e., comparing the project with other puzzles or game consoles). The author is planning to publish the results in a separate paper.

# Appendix 1.

**The technical description of the assembled prototype.**

The prototype consists of 8 structurally identical sub-elements and one central element.

Each sub element consists of a shell, with three screens, and three output pin ports
The following components were used:

1. TP4056 Precise 5V Micro USB Lithium Battery Charging Power

2. Serial LCD Display 128*128 SPI TFT Color Screen With PCB Adapter

3. JST 2.5 XH 8-Pin Connector plug with Wire

4. DC-DC Boost Converter Step Up Module 1-5V to 5V 500mA Power Module

5. Arduino pro mini  Microcontroller 2016 Enhancement 5V adjustable 16M MEGA328P

6. 3.5mm Male AUX Audio Jack To USB 2.0 Male Charge Cable Adapter Cord Black

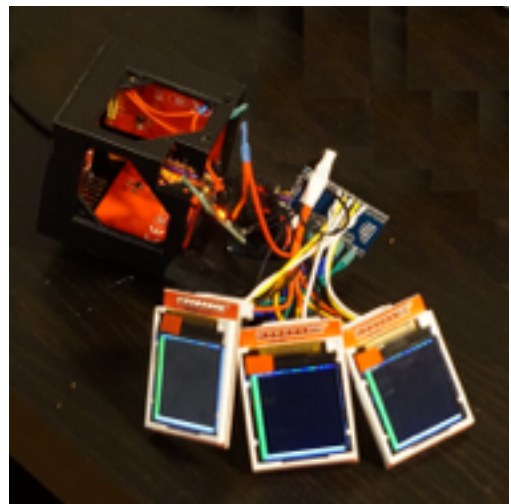

7. Phillips Flat Countersunk Head Self Tapping Screw m2x6mm for Most Remote

8. Li-ion Rechargeable Battery 2400mAh  3.7V 16340 CR123A

9. Switch On-Off Black Mini Size SPDT Slide PCB DIY 5V 0.3A Compact & Durable

10  STRONG MAGNETS 6mm (7/32") spheres balls N35 Neodymium - 100

11 BLACK PLA 3D Printer Filament 1.75mm for 3D-Print RepRap Rapman

12 TRRS 3.5mm 4-Conductor Chassis Mount Snap-in Jack-Black

The central element. It can be in two versions:

Preciball Chrome Steel Ball  Fist Self Defense Craft Heavy Weights 1"
OR
2x2 Cube Puzzle 1"

**Source code of the software in the c++ language**
by the address https://github.com/ilya000/cubios_001

**Electric schematic diagram**

(For each sub-element)

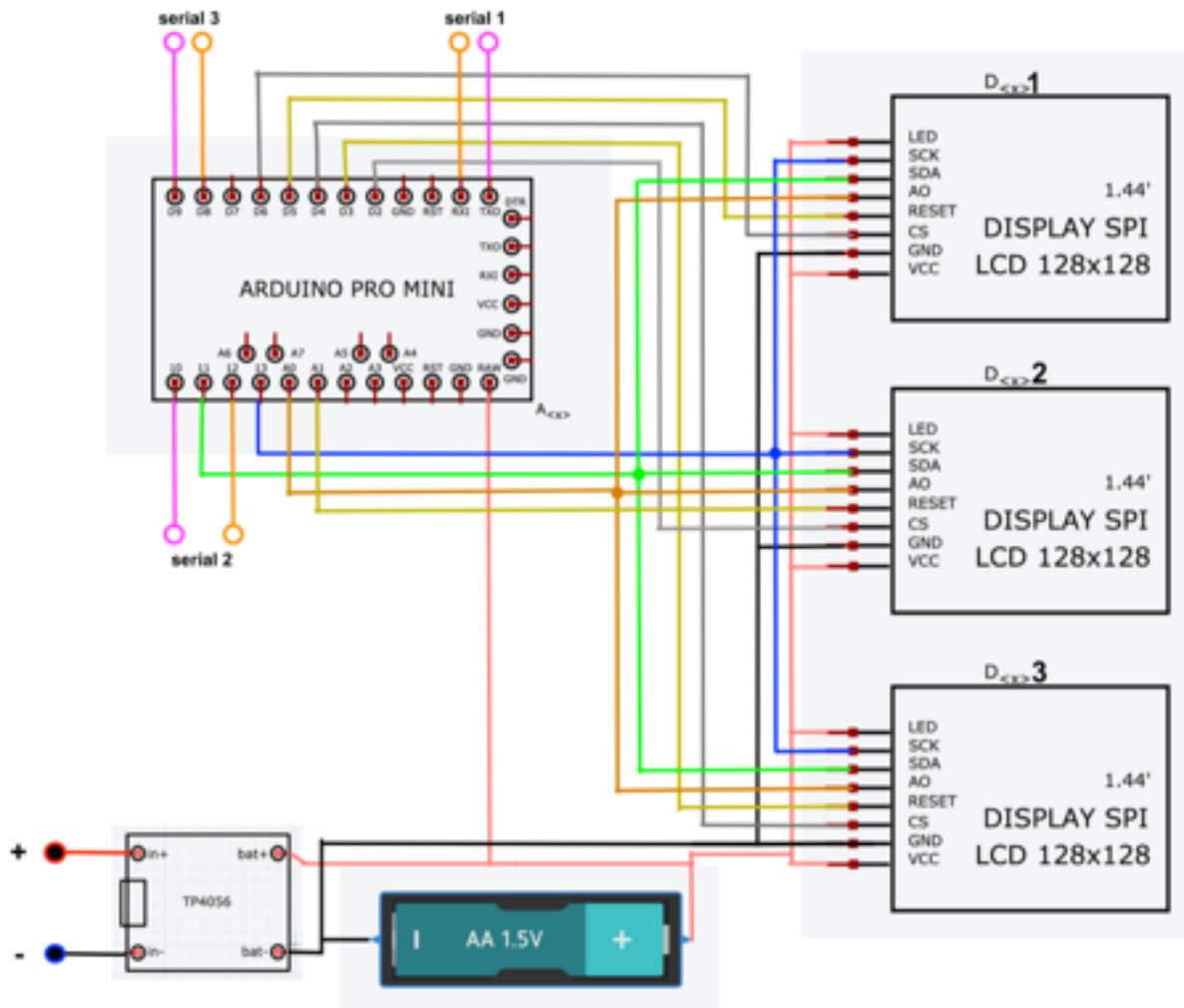

**Plastic elements printed on a 3D printer**
(For each sub-element)

1. Housing

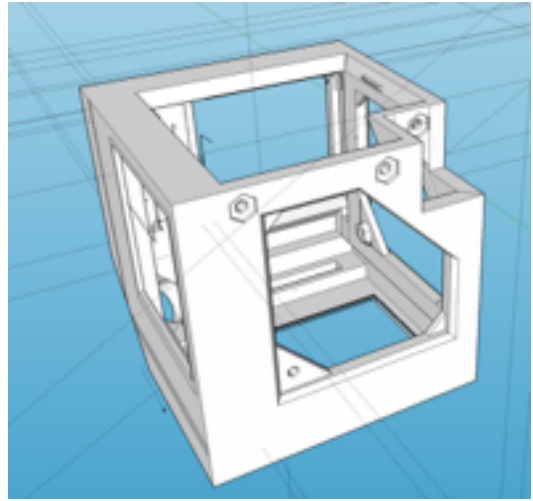

2. Lid type 1

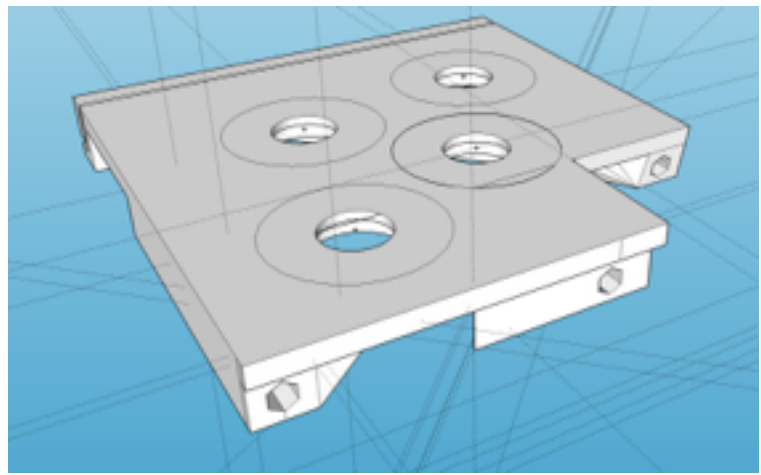

3. Lid of type 2, and a magnet holder

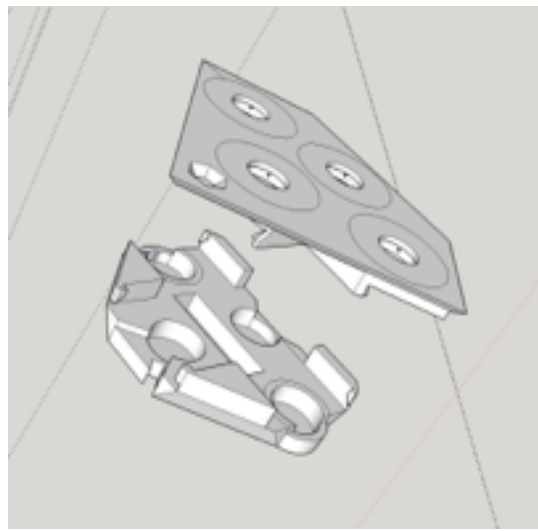